\documentclass[conference,onecolumn]{IEEEtran}
\IEEEoverridecommandlockouts
\usepackage{siunitx}
\usepackage[acronym]{glossaries}
\usepackage[sorting=none]{biblatex}
\usepackage{ifthen}
\usepackage{amsmath,amssymb,amsfonts}
\usepackage{algorithmic}
\usepackage{graphicx}
\usepackage{textcomp}
\usepackage{xcolor}
\usepackage{colortbl}
\usepackage{ragged2e}
\usepackage{tabularx}
\usepackage[bookmarks=false,hidelinks=true]{hyperref}
\usepackage{comment}
\usepackage{booktabs}
\usepackage{multirow}
\usepackage{dirtytalk}
\usepackage{tikz}
\usetikzlibrary{shapes.geometric, arrows, positioning}
\usepackage{circuitikz}
\usepackage{svg}
\usepackage{pgfplots}
\usepgfplotslibrary{fillbetween}
\pgfplotsset{width=10cm,compat=1.9}
\usepackage[most]{tcolorbox}
\usepackage{enumitem}
\usepackage{caption}
\usepackage{listings}
\usepackage{float}

\definecolor{ForestGreen}{RGB}{34,150,34}

\newboolean{blindreview}
\setboolean{blindreview}{false}
\DeclareRobustCommand{\IEEEauthorrefmark}[1]{\smash{\textsuperscript{\footnotesize #1}}}

\newacronym{SoC}{SoC}{System-on-Chip}
\newacronym{CWEs}{CWEs}{Common Weakness Enumerations}
\newacronym{CWE}{CWE}{Common Weakness Enumeration}
\newacronym{CVEs}{CVEs}{Common Vulnerability Enumerations}
\newacronym{CVE}{CVE}{Common Vulnerability Enumeration}
\newacronym{AI}{AI}{Artificial Intelligence}
\newacronym{ML}{ML}{Machine Learning}
\newacronym{LLMs}{LLMs}{Large Language Models}
\newacronym{LLM}{LLM}{Large Language Model}
\newacronym{ASIC}{ASIC}{Application Specific Integrated Circuit}
\newacronym{DUV}{DUV}{Design Under Verification}
\newacronym{CEX}{CEX}{Counter Example}
\newacronym{CEXs}{CEXs}{Counter Examples}
\newacronym{DOS}{DOS}{Denial of Service}
\newacronym{FSM}{FSM}{Finite State Machine}
\newacronym{FSMs}{FSMs}{Finite State Machines}
\newacronym{SEU}{SEU}{Single Event Upset}
\newacronym{SEUs}{SEUs}{Single Event Upsets}
\newacronym{RTL}{RTL}{Register Transfer Level}
\newacronym{IP}{IP}{Intellectual Property}
\newacronym{NLP}{NLP}{Natural Language Processing}
\newacronym{HDL}{HDL}{Hardware Description Language}
\newacronym{GPT}{GPT}{Generative Pre-trained Transformer}
\newacronym{PPA}{PPA}{Power, Performance and Area}
\newacronym{SVA}{SVA}{SystemVerilog Assertion}
\newacronym{SVAs}{SVAs}{SystemVerilog Assertions}
\newacronym{FV}{FV}{Formal Verification}
\newacronym{CSV}{CSV}{Comma-Separated Values}
\newacronym{RQ}{RQ}{Research Question}
\newacronym{RQs}{RQs}{Research Questions}
\newacronym{GenAI}{GenAI}{Generative AI}
\newacronym{UVM}{UVM}{Universal Verification Methodology}
\newacronym{ADHD}{ADHD}{Attention Deficit Hyperactivity Disorder}
\newacronym{LRM}{LRM}{Language Reference Manual}
\newacronym{PoC}{PoC}{Proof of Concept}

\begin{document}

\lstset{
    language=Verilog,           
    basicstyle=\footnotesize,   
    numbers=left,               
    frame=lines,                
    captionpos=b,               
    breaklines=true,            
    tabsize=2,                  
    xleftmargin=2.1em,
    framexleftmargin=1.7em,
    commentstyle=\color{ForestGreen},
    keywordstyle=\color{blue},
    stringstyle=\color{red},
}

\lstdefinelanguage{Verilog}{morekeywords={accept_on,alias,always,always_comb,always_ff,always_latch,and,assert,assign,assume,automatic,before,begin,bind,bins,binsof,bit,break,buf,bufif0,bufif1,byte,case,casex,casez,cell,chandle,checker,class,clocking,cmos,config,const,constraint,context,continue,cover,covergroup,coverpoint,cross,deassign,default,defparam,design,disable,dist,do,edge,else,end,endcase,endchecker,endclass,endclocking,endconfig,endfunction,endgenerate,endgroup,endinterface,endmodule,endpackage,endprimitive,endprogram,endproperty,endspecify,endsequence,endtable,endtask,enum,event,eventually,expect,export,extends,extern,final,first_match,for,force,foreach,forever,fork,forkjoin,function,generate,genvar,global,highz0,highz1,if,iff,ifnone,ignore_bins,illegal_bins,implements,implies,import,incdir,include,initial,inout,input,inside,instance,int,integer,interconnect,interface,intersect,join,join_any,join_none,large,let,liblist,library,local,localparam,logic,longint,macromodule,matches,medium,modport,module,nand,negedge,nettype,new,nexttime,nmos,nor,noshowcancelled,not,notif0,notif1,null,or,output,package,packed,parameter,pmos,posedge,primitive,priority,program,property,protected,pull0,pull1,pulldown,pullup,pulsestyle_ondetect,pulsestyle_onevent,pure,rand,randc,randcase,randsequence,rcmos,real,realtime,ref,reg,reject_on,release,repeat,restrict,return,rnmos,rpmos,rtran,rtranif0,rtranif1,s_always,s_eventually,s_nexttime,s_until,s_until_with,scalared,sequence,shortint,shortreal,showcancelled,signed,small,soft,solve,specify,specparam,static,string,strong,strong0,strong1,struct,super,supply0,supply1,sync_accept_on,sync_reject_on,table,tagged,task,this,throughout,time,timeprecision,timeunit,tran,tranif0,tranif1,tri,tri0,tri1,triand,trior,trireg,type,typedef,union,unique,unique0,unsigned,until,until_with,untyped,use,uwire,var,vectored,virtual,void,wait,wait_order,wand,weak,weak0,weak1,while,wildcard,wire,with,within,wor,xnor,xor,`uvm_create, `uvm_rand_send_with},morecomment=[l]{//}}

\title{ConnChecker: Automated Root-Cause Analysis for Formal Connectivity Check via Graph \\
}

\ifthenelse{\boolean{blindreview}}{}{
    \author{
        \IEEEauthorblockN{
            Do Ngoc Tiep\IEEEauthorrefmark{1},
            Nguyen Linh Anh\IEEEauthorrefmark{2},
            Luu Danh Minh\IEEEauthorrefmark{1}
        }
        \IEEEauthorblockA{
            \IEEEauthorrefmark{1}Infineon Technologies Vietnam Company Ltd., Hanoi, Vietnam\\
            \IEEEauthorrefmark{2}Cornell University, Ithaca, New York, USA
        }
        \IEEEauthorblockA{
            \textit{E-mail: ngoctiep.do@infineon.com, lan64@cornell.edu, danhminh.luu@infineon.com}
        }
    }
}

\maketitle
\begin{center}
{\textit{Dedication}} \\[0.1cm]
\textit{To our Functional Manager, who has always supported us and placed unwavering trust in our work.}
\end{center}
\vspace{0.2cm}

\begin{abstract}
\textbf{Formal connectivity checking offers scalable verification of signal paths in complex SoC designs, but debugging counterexamples remains a manual and time-consuming process. ConnChecker introduces a new graph-based perspective for automating root-cause analysis by integrating formal tool outputs such as structural/functional dependency graphs and counterexamples report. It begins with automatic failure categorization, routing each counterexample to one of three targeted analysis flows. These flows localize failure points and suggest corrective actions or hints for manual inspection. Evaluated on two industrial SoCs, ConnChecker achieved up to 80\% reduction in debugging time, especially for complex cases, demonstrating its scalability and effectiveness across diverse connectivity scenarios.}
\end{abstract}


\section{Introduction} \label{intro}

While formal verification provides high confidence in hardware correctness, its scalability at the SoC level is often constrained by state-space explosion. A notable exception is connectivity checking, which verifies signal paths and interconnections with tractable complexity. Standard workflows involve building a connectivity map from design specifications and using formal tools, such as the Cadence JasperGold Connectivity App~\cite{cadence_jasper_conn_app}, for verification. When failures occur, counterexamples (CEXs) are generated, requiring manual and error-prone debugging. This process is among the most time-consuming tasks in both formal and simulation-based functional verification. According to the 2024 Wilson Research Group IC/ASIC Functional Verification Trend Report~\cite{foster2024verification}, engineers spend approximately 47\% of their time on debugging. As design complexity increases, efficient debugging methodologies are urgently needed. Although prior work has explored formal connectivity checking~\cite{7843024, 9119174}, to our knowledge, no publicly available solution focuses on accelerating the debugging process.

To address this gap, we propose ConnChecker, a graph-based framework designed to accelerate root-cause analysis (RCA) in formal connectivity verification. ConnChecker processes CEX reports along with the functional and structural dependency graph generated by JasperGold, categorizing failure types and applying targeted analysis workflows. These workflows help identify root causes, suggest potential fixes, or narrow down the search space for manual inspection, significantly reducing debugging effort.

\begin{figure}[!h]
    \centering
    \includegraphics[width=0.85\linewidth]{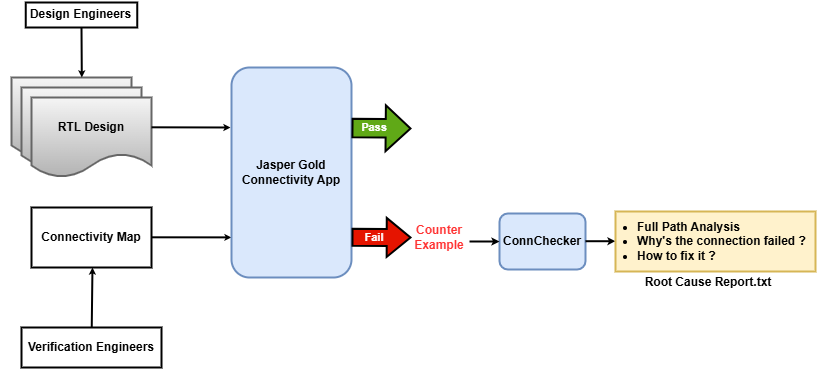}
    \caption{ConnChecker to accelerate connectivity debug process}
    \label{fig:connchecker-intro}
\end{figure}

ConnChecker reduces manual effort and enhances diagnostics through three key innovations:

\begin{enumerate}
    \item \textbf{New perspective on automating connectivity checking:} ConnChecker introduces a structured failure categorization stage that automatically assigns each counterexample to one of three analysis flows, transforming manual debugging into an automated, scalable, and graph-driven process.
    \item \textbf{Automated generation of atomic connectivity checks:} For each segment in the dependency graph, ConnChecker generates targeted assertions to isolate broken edges, enabling precise root-cause localization with minimal manual intervention.
    \item \textbf{Fan-in analysis of the destination signal’s cone of influence:} ConnChecker traces upstream dependencies to identify missing drivers and disconnected segments, streamlining structural debugging in cases with no connectivity.
\end{enumerate}

We evaluated ConnChecker on two commercial SoC designs, a radar sensor~\cite{infineon_60ghz_radar} and an automotive microcontroller~\cite{infineon_aurix}, to validate its feasibility and effectiveness. Across three distinct analysis flows, ConnChecker consistently reduced debugging time compared to manual methods. For simple connections, performance was comparable. However, as complexity increased, ConnChecker achieved up to 80\% time savings. These results demonstrate strong scalability and practical value of ConnChecker in accelerating formal connectivity debugging.

\section{Problem Formulation} \label{problem_formulation}
  
The design-under-test (DUT) is modeled as a directed dependency graph \( G = (V, E) \), where each node, formally vertex, \( v \in V \) represents a signal, pin, or variable, and each edge \( e = (v_i, v_j) \in E \) denotes a direct dependency or structural connection between nodes. Hierarchical graphs from individual IP blocks are merged to form subsystem-level graphs, and ultimately a system-level graph representing the full SoC.

Given a source node \( v_s \in V \) and a destination node \( v_d \in V \), a connectivity check asserts the existence of a valid path \( P = \{v_s, v_1, \ldots, v_d\} \subseteq G \) such that signal propagation from \( v_s \) to \( v_d \) is feasible under the design’s structural and functional constraints.

\begin{figure}[!h]
    \centering
    \includegraphics[width=0.85\linewidth]{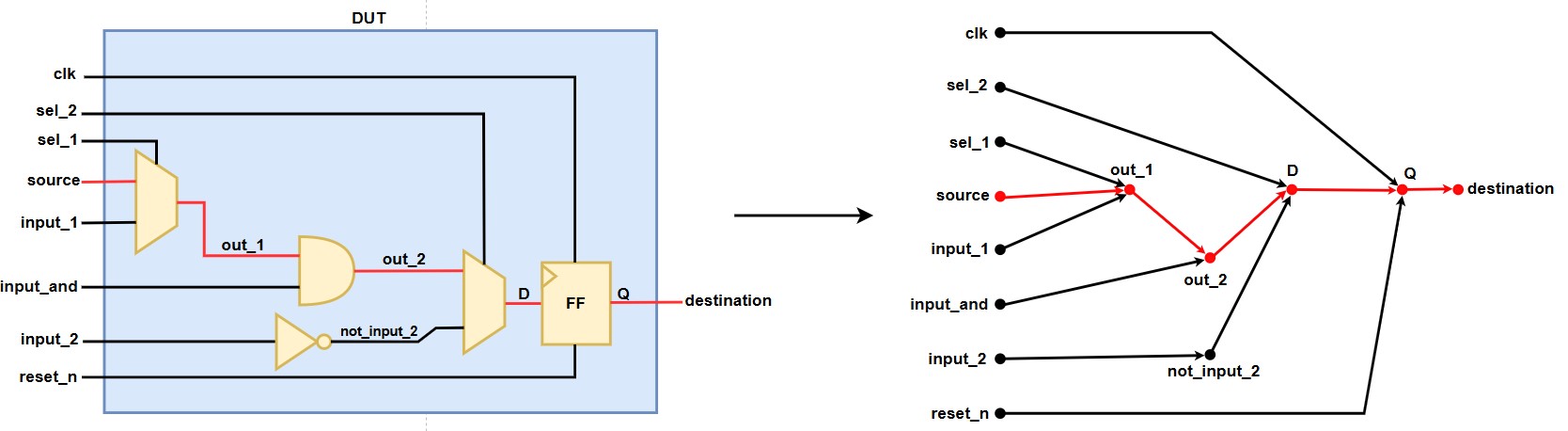}
    \caption{A simple RTL design modeled as a directed dependency graph \(G = (V, E)\). To check connectivity from \textit{source} (\(v_s\)) to \textit{destination} (\(v_d\)), we consider a path \(P \subseteq G\), where \(P = \{v_s, \text{out\_1}, \text{out\_2}, D, Q, v_d\}\) and the corresponding edges are \(E_P = \{(v_s, \text{out\_1}), (\text{out\_1}, \text{out\_2}), (\text{out\_2}, D), (D, Q), (Q, v_d)\}\).}
    \label{fig:circuit_to_graph}
\end{figure}

JasperGold Connectivity App defines two types of connections: structural and functional~\cite{cadence_jasper_conn_app}. A structural connection refers to an RTL-defined path between two signals, visible in the cone of influence, regardless of whether signal propagation is possible. A functional connection is a subset of structural connections where the driving signal can logically influence the receiving signal, considering design constraints and gating logic. These definitions form the foundation for classifying connectivity failures into three distinct categories:

\begin{enumerate}[label=\arabic*.]
    \item \textbf{Functional and structural connection exists:}  
    The path \(P\) is structurally present and functionally valid, but the check fails due to issues such as incorrect connectivity maps, RTL bugs, or missing constraints.

    \item \textbf{No structural connection:}  
    One or more edges in \(E_P\) are missing (e.g., \((out\_1, out\_2)\)), making \(v_d\) unreachable from \(v_s\), indicating a structural issue.

    \item \textbf{Only structural connection exists:}  
    The path \(P\) exists structurally, but signal propagation is blocked due to over-constraints (e.g., \(\textit{input\_and is tied to 0}\)), disabled logic, or incomplete formal setup, resulting in a functional disconnect.
\end{enumerate}

The goal is to develop a streamlined solution that quickly classifies each failure and triggers the appropriate analysis workflow, enabling efficient root-cause analysis and reducing debugging effort.

\section{Proposed Solution} \label{proposed_solution}

In this section, we describe our two-stage solution to enhance JasperGold Connectivity App based on our goal.

\subsection{Failure Categorization Flow}

\begin{figure}[!h]
    \centering
    \includegraphics[width=0.6\linewidth]{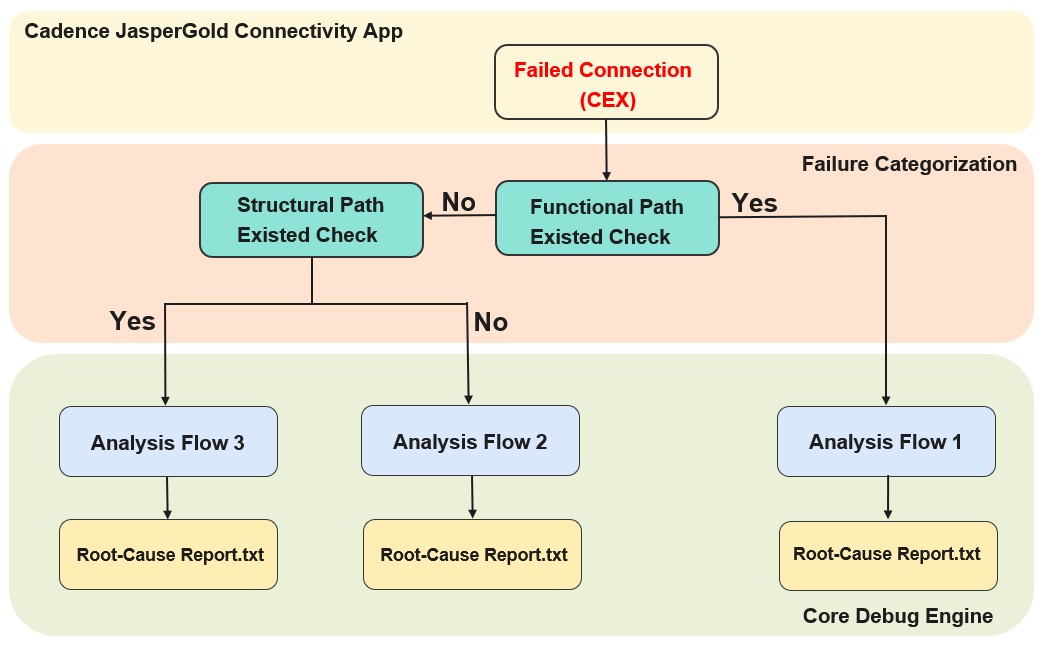}
    \caption{Failure categorization flow}
    \label{fig:failure-classification-flow}
\end{figure}

The first stage is an automatic failure categorization flow that maps each CEX to its failure type. Based on the pre-defined failure categories from the previous section, we implement a lightweight yet effective decision flow, illustrated in Figure~\ref{fig:failure-classification-flow}. Starting from the CEX, ConnChecker invokes JasperGold to generate the functional path from source to destination. If a valid functional path is found, the failure is routed to Analysis Flow 1. Otherwise, ConnChecker checks for a structural path to determine whether the failure is assigned to Analysis Flow 3 for structural-only or Analysis Flow 2 for no structural connection.

\subsection{Core Debug Engine}

Once the right category is identified, a CEX is funneled into one of the three Analysis Flows within the core debug engine:

\subsubsection{\textbf{Analysis Flow 1 - Functional and structural connection exists}}

\begin{figure}[H]
    \centering
    \includegraphics[width=1\linewidth]{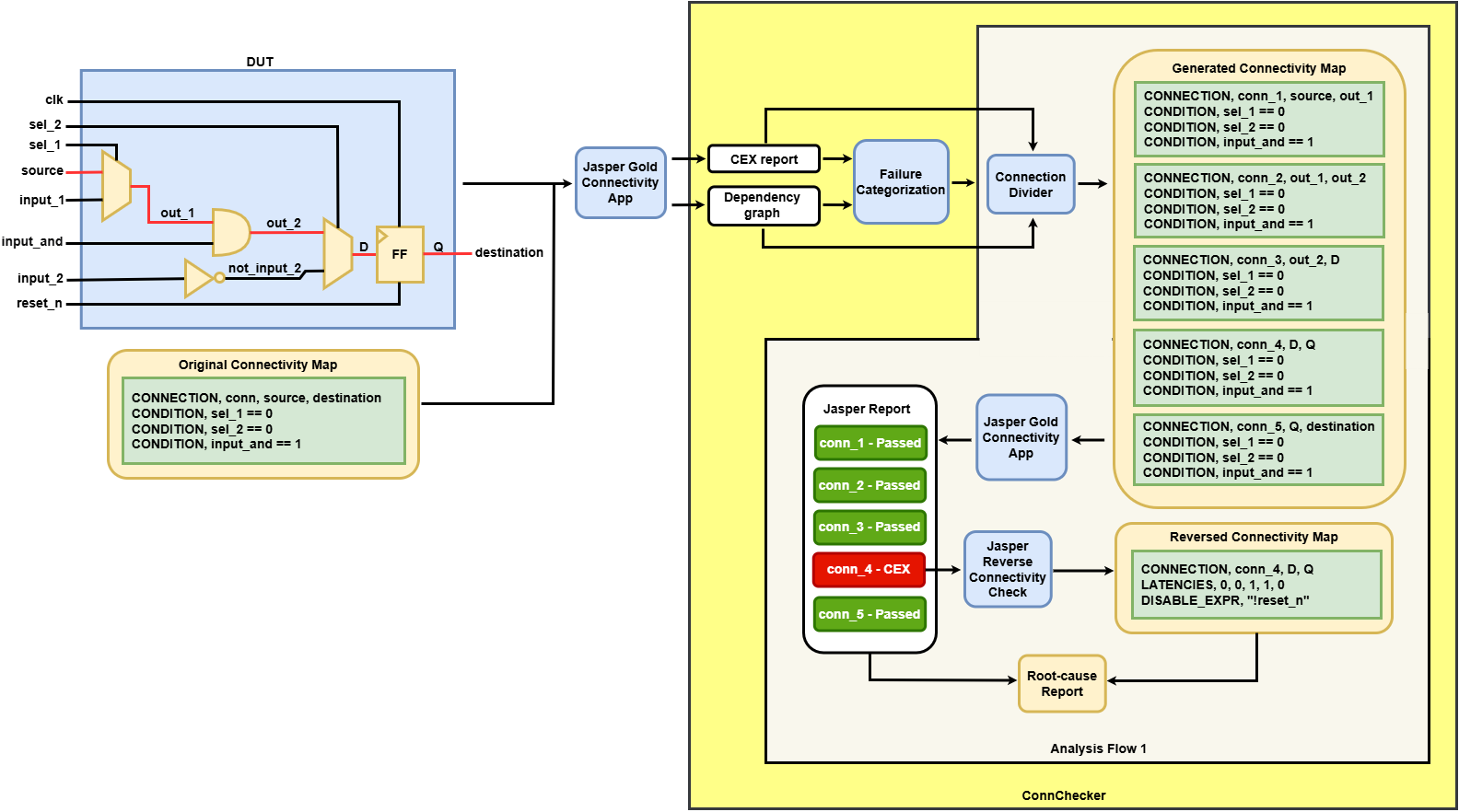}
    \caption{Analysis Flow 1 - Segment-level connectivity for identifying the root-cause on the path}
    \label{fig:analysis-flow-one}
\end{figure}

This flow, depicted in Figure~\ref{fig:analysis-flow-one}, addresses cases where a functional path is expected but the connectivity check fails due to issues such as incorrect connectivity maps, RTL bugs, or missing constraints. The analysis engine examines the functional dependency graph generated by JasperGold, decomposing it into smaller segments for detailed inspection. Furthermore, ConnChecker extracts connection names and conditions from the CEX report. For each segment, ConnChecker combines the segment with its extracted data to build a refined segment-level connectivity map, which is re-verified with JasperGold to identify the failing segment(s). In essence, ConnChecker generates atomic assertions for each edge of the dependency graph to confirm functional connectivity. This enables precise localization of broken edge(s) along the full path.

To further enhance the analysis, JasperGold's reverse connectivity check is applied to failed segments. This technique uses formal methods to extract connections directly from the RTL, allowing users to validate them or compare them against the specification. ConnCheck compiles the results into a unified report that outlines the full path structure, pinpoints failure causes, and recommends corrective actions. While applying the reverse check to the entire path may seem ideal, practical limitations make segment-level analysis more effective. First, segmenting the path provides finer granularity, offering deeper insight into each component-to-component connection. Second, applying reverse connectivity check on each segment gives a clear picture, allowing engineers to know the exact conditions required for each segment. Third, the runtime of the reverse connectivity check for the full path is much longer than for failed segment(s) only, making segment-level analysis more efficient in practice. Most importantly, reverse connectivity checks rely on path depth as input, which varies across designs and complicates automation. In contrast, segment-level checks, always have the depth of one, enabling us to create a simpler, fully automated flow that meets the needs across various designs.

\subsubsection{\textbf{Analysis Flow 2 - No structural connection}}

\begin{figure}[h]
    \centering
    \includegraphics[width=1\linewidth]{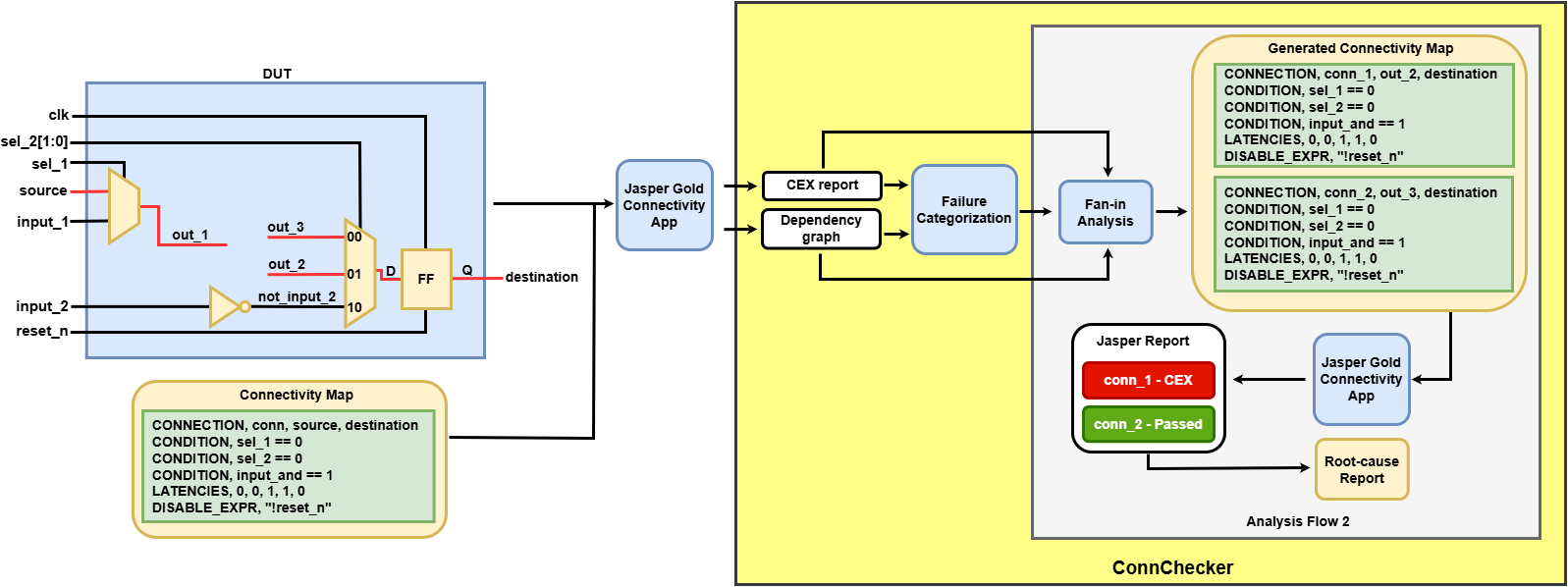}
    \caption{Analysis Flow 2 for identifying breakpoint(s) during connectivity check}
    \label{fig:analysis-flow-two}
\end{figure}

This flow, illustrated in Figure~\ref{fig:analysis-flow-two}, addresses failures caused by missing structural connections between the source and destination signals. When no direct path exists, traditional connectivity checks cannot proceed. To handle this, we use fan-in analysis, a backward-tracing method that starts from the destination signal and identifies all signals that directly or indirectly influence it. The goal is to build a fan-in graph that reveals the destination's cone of influence and highlights missing drivers or disconnected segments.

The fan-in graph differs from a full structural dependency graph by focusing solely on signals that influence the destination, rather than capturing the entire connection. JasperGold displays all signals, both directly and indirectly, that affect the destination signal. ConnChecker enhances this by instructing JasperGold to filter out irrelevant signals such as clocks, resets, and top-level inputs. It then identifies root drivers, signals that are not driven by any others and may indicate potential breakpoints, before constructing the fan-in graph. Using this refined output, ConnChecker generates a connectivity map to trace paths from the root drivers to the destination. This map reuses the original connection conditions, so if a root driver’s connection check passes, it suggests that the driver could be the breakpoint. ConnChecker reports all identified root drivers to engineers, offering a clear and concise set of candidates for further debugging. By combining targeted insights with automated analysis, ConnChecker streamlines the debugging process and provides actionable results, significantly improving upon manual inspection of raw tool outputs.

\begin{figure}[h]
    \centering
    \includegraphics[width=1\linewidth]{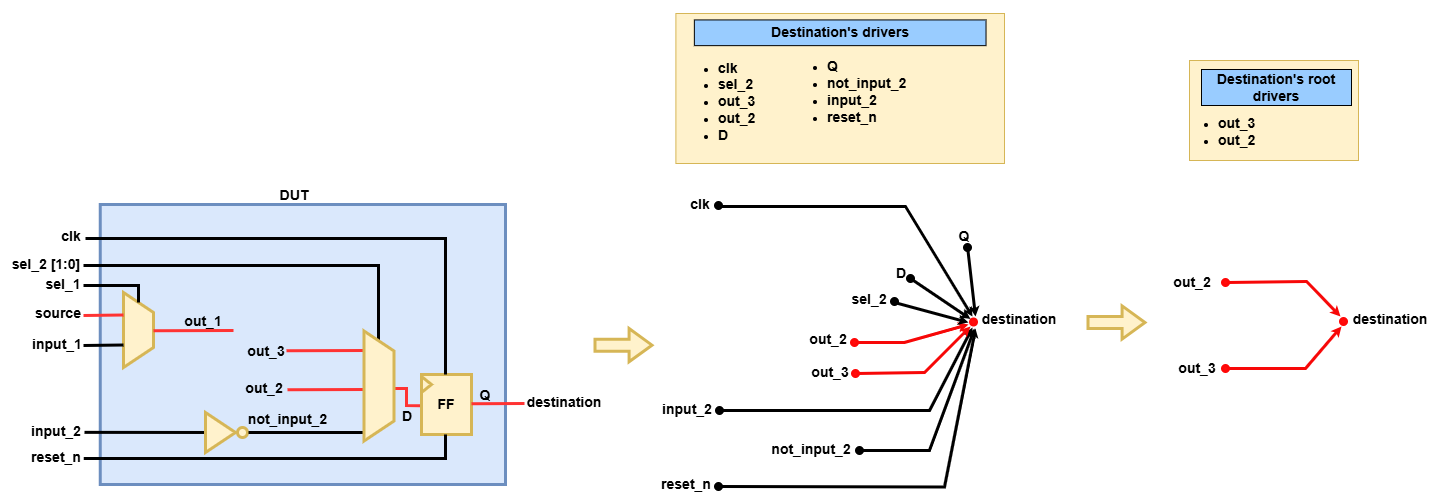}
    \caption{Potential destination's root drivers \(\text{out\_2}\) and \(\text{out\_3}\) in the circuit reflected in the fan-in graph}
    \label{fig:analysis-flow-two-example}
\end{figure}

Take the example in Figure~\ref{fig:analysis-flow-two-example}. JasperGold generates the destination's drivers list:
\[
\{\texttt{clk}, D, Q, \texttt{sel\_2}, \texttt{out\_2},  \texttt{out\_3}, \texttt{not\_input\_2}, \texttt{input\_2}, \texttt{reset\_n}\}
\]
Excluding clock, reset, selection, and top-level input signals, only \texttt{out\_2} and \texttt{out\_3} remain as drivers that are not driven by any other signal. We refer these as the root drivers of the destination.

\subsubsection{\textbf{Analysis Flow 3 - Only structural connection exists}} 
    
This flow targets scenarios where a structural path exists, but no functional propagation is observed, typically due to over-constraints, RTL logic issues, or incomplete formal environments. ConnChecker applies a divide-and-conquer strategy, similar to Analysis Flow 1, to the structural dependency path. While reverse connectivity checks are not applicable to over-constrained segments, JasperGold offers alternative techniques such as over-constraint analysis~\cite{cadence_jasper_conn_app}, which ConnChecker leverages to effectively pinpoint and explain the root causes of functional disconnects. 

\subsection{Single Mode and Multi Mode}
 To support a range of debugging scenarios, ConnChecker provides two operational modes: \textbf{single mode}, designed for analyzing individual failures, and \textbf{multi mode}, which enables concurrent processing of multiple failures for scalable, high-throughput debugging. In this work, we focus on evaluating the effectiveness of the single mode to highlight the core capabilities of ConnChecker.

\section{Experimental setup, results and Discussion} \label{preliminary_results}

\subsection{Experimental Setup}
We evaluated ConnChecker on two commercial SoC designs: a radar sensor from the Infineon 60GHz RADAR SENSOR family~\cite{infineon_60ghz_radar} and an automotive microcontroller from the Infineon AURIX MCU family~\cite{infineon_aurix}. The radar sensor is medium-sized, integrating compute and control subsystems with standard peripheral interfaces for configuration, data movement, and design-for-test (DFT). Its connectivity spans multiple hierarchical levels, with thousands of top-level and intra-subsystem connections across buses, point-to-point signals, and pad/IO wiring. The microcontroller shares these characteristics but at a significantly larger scale, with more modules, deeper hierarchy, and broader peripheral coverage. Both designs are production-grade, taped-out chips selected to cover a wide range of connectivity scenarios and validate ConnChecker’s applicability in real-world industrial environments.

\begin{figure}[H]
    \centering
    \includegraphics[width=0.8\linewidth]{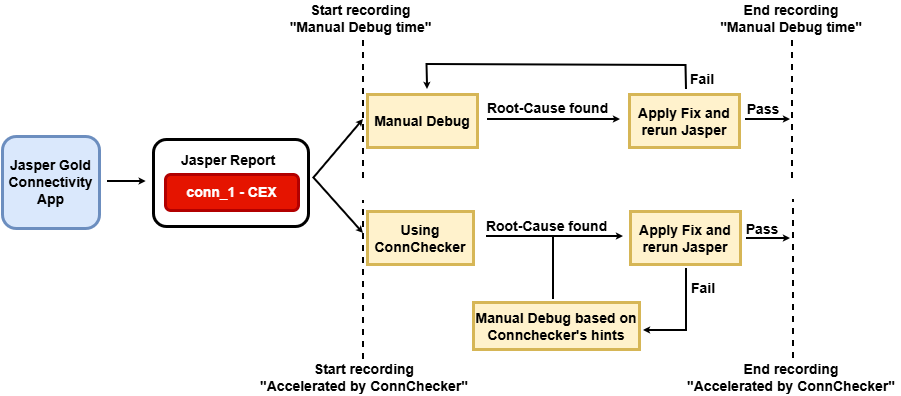}
    \caption{Experimental Setup}
    \label{fig:experimental-setup}
\end{figure}

To evaluate the tool's effectiveness, we applied ConnChecker's three Analysis Flows to a curated set of single-failure cases across both designs. Each case varies in complexity, defined by the number of graph edges and the nature of the connectivity path such as combinational, sequential, mixed-signal, or multi-clock domain. This setup enables a structured assessment of the graph-based approach and its ability to handle varying connectivity characteristics. In addition, we also compared the time required for manual debugging against the time accelerated by ConnChecker with the process shown in Figure~\ref{fig:experimental-setup}. The comparison offers a quantitative measure of productivity gains and demonstrates the practical benefits of automation in reducing debug turnaround time. Detailed results are presented in the following section.

\subsection{Experimental Results}

\begin{figure} [H]
    \centering
    \includegraphics[width=0.6\linewidth]{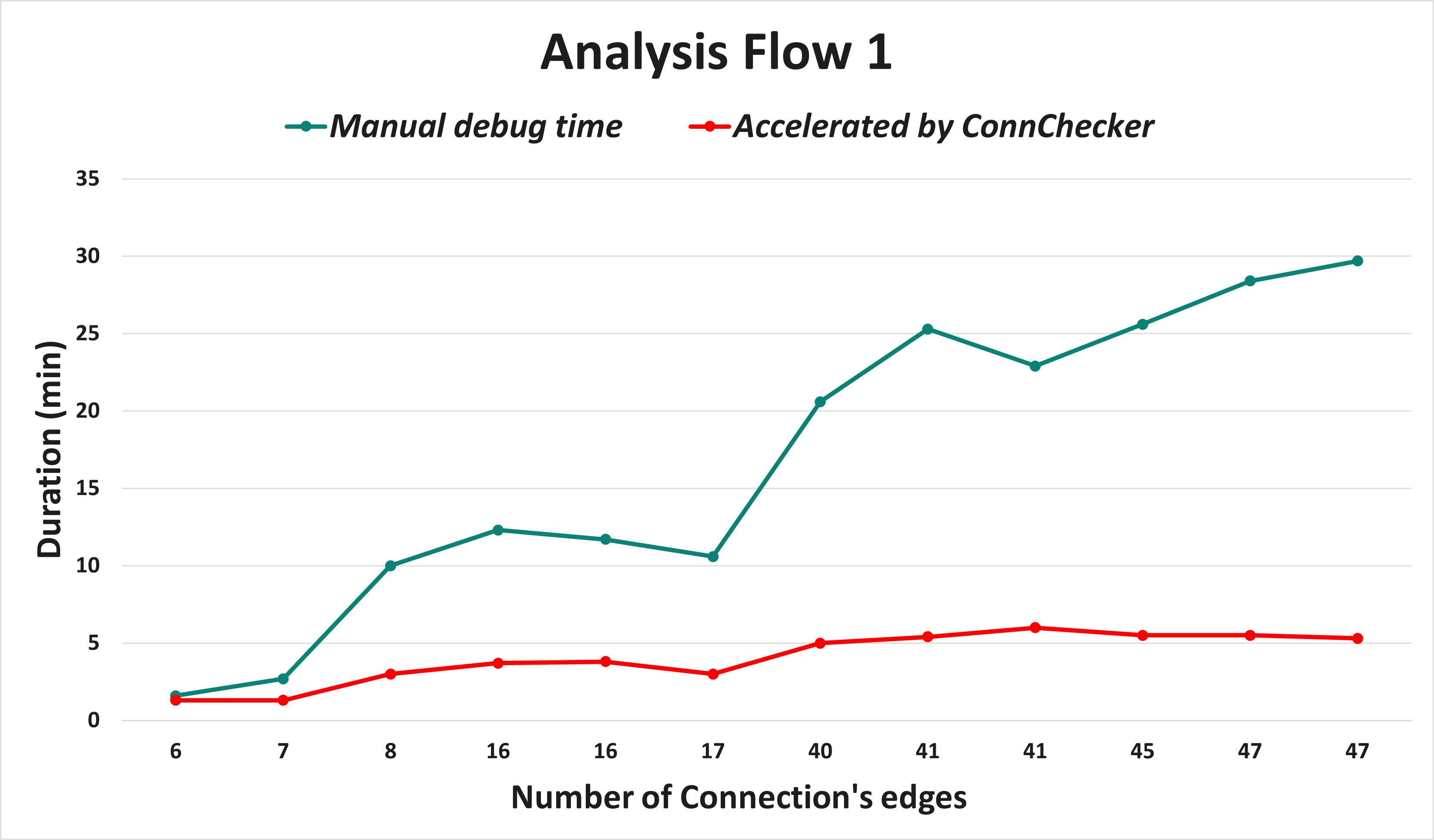}
    \caption{Graphical comparison of Table~\ref{tab:flow-1-table}.}
    \label{fig:flow-1-graph}
\end{figure}

Table~\ref{tab:flow-1-table} and Figure~\ref{fig:flow-1-graph} present the runtime comparison between manual debugging and ConnChecker for cases under Analysis Flow 1. ConnChecker achieves a consistent reduction in debugging time, with speedups ranging from 2× to over 5× depending on connection complexity. In high-complexity cases involving mixed-signal and multi-clock domains (e.g., Case 12), ConnChecker reduces analysis time from nearly 30 minutes to approximately 5 minutes. This improvement is enabled by automated segment-level decomposition, targeted assertion generation, and integration with reverse connectivity checks.

\begin{table}[H]
  \centering
  \footnotesize
  \setlength{\tabcolsep}{2pt}
  \renewcommand{\arraystretch}{1.1}
  \begin{tabular}{|>{\centering\arraybackslash}p{1.5cm}|p{2.5cm}|p{7.0cm}|>{\centering\arraybackslash}p{1.5cm}|>{\centering\arraybackslash}p{2.0cm}|>{\centering\arraybackslash}p{2.5cm}|}
    \hline
    \textbf{Connection} & \textbf{Design type} & \textbf{Connection's characteristic} & \textbf{No. of connection's edges} & \textbf{Manual debug time (HH:MM:SS)} & \textbf{Accelerated by ConnChecker (HH:MM:SS)} \\
    \hline \hline
    1  & Radar Sensor                 & Combinational                                             & 7  & 00:02:39 & 00:01:15  \\
    2  & Radar Sensor                 & Combinational                                             & 6  & 00:01:33 & 00:01:20  \\
    3  & Radar Sensor                 & Combinational/Mixed-signal                                & 8  & 00:10:02 & 00:03:01  \\
    4  & Radar Sensor                 & Combinational/Mixed-signal                                & 16 & 00:12:15 & 00:03:40  \\
    5  & Radar Sensor                 & Combinational/Mixed-signal                                & 16 & 00:11:40 & 00:03:45  \\
    6  & Radar Sensor                 & Combinational/Mixed-signal                                & 17 & 00:10:34 & 00:02:59  \\
    7  & ATV Microcontroller          & Combinational/Mixed-signal                                & 41 & 00:22:20 & 00:05:23  \\
    8  & ATV Microcontroller          & Combinational/Mixed-signal                                & 40 & 00:20:34 & 00:05:01  \\
    9  & ATV Microcontroller          & Combinational/Mixed-signal                                & 41 & 00:22:53 & 00:05:59  \\
    10 & ATV Microcontroller          & Combinational/Sequential/Mixed-signal                     & 45 & 00:25:34 & 00:05:32  \\
    11 & ATV Microcontroller          & Combinational/Sequential/Mixed-signal/Multi-clock domain  & 47 & 00:28:26 & 00:05:29  \\
    12 & ATV Microcontroller          & Combinational/Sequential/Mixed-signal/Multi-clock domain  & 47 & 00:29:44 & 00:05:20  \\
    \hline
  \end{tabular}
  \caption{Analysis Flow 1 - Cases with both functional and structural connections.}
  \label{tab:flow-1-table}
\end{table}

\begin{figure} [H]
    \centering
    \includegraphics[width=0.6\linewidth]{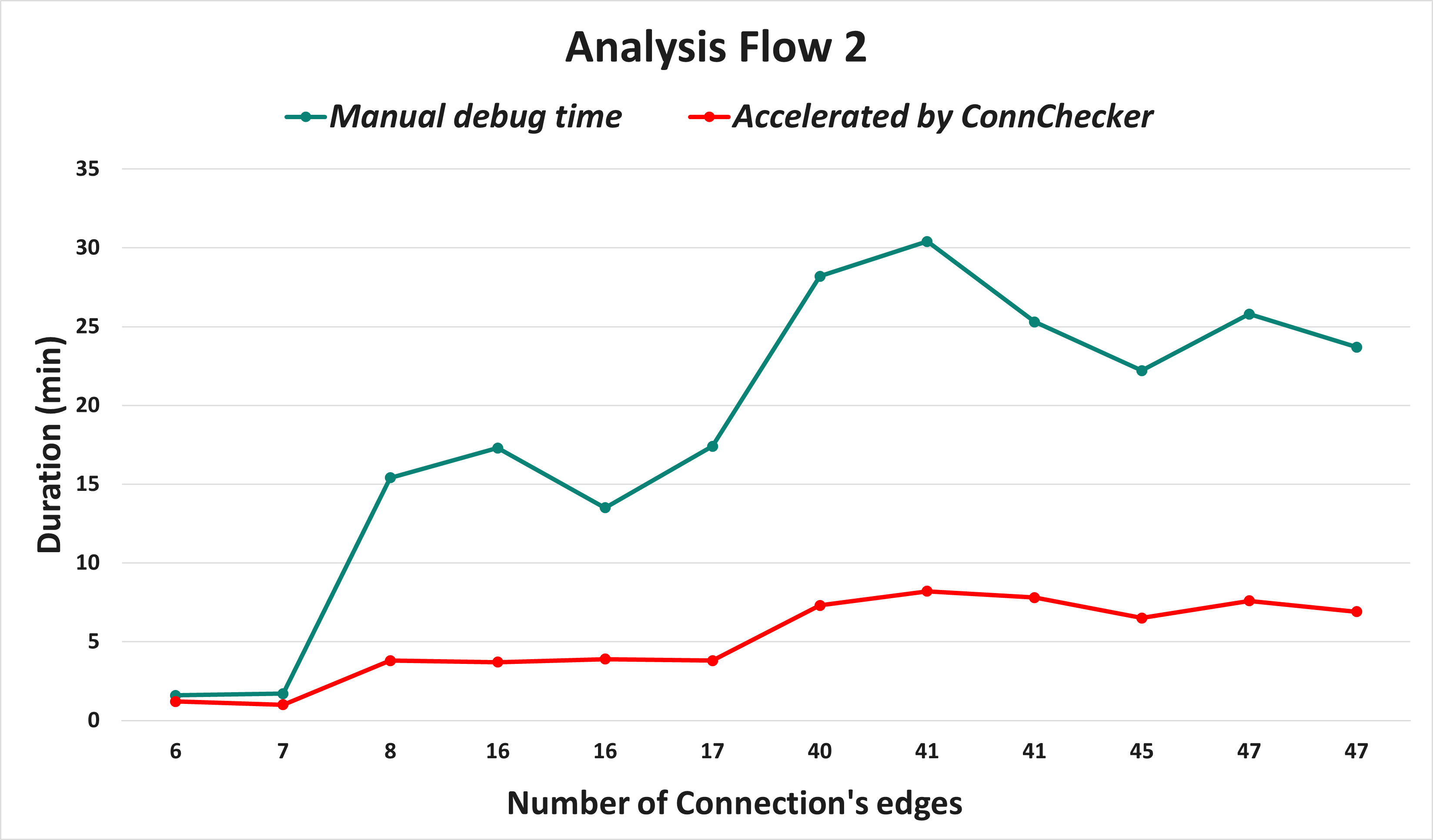}
    \caption{Graphical comparison of Table~\ref{tab:flow-2-table}.}
    \label{fig:flow-2-graph}
\end{figure}

Table~\ref{tab:flow-2-table} and Figure~\ref{fig:flow-2-graph} showcase ConnChecker’s performance in Analysis Flow 2. Similar to the Analysis Flow 1, ConnChecker significantly reduces debugging time across all cases, with speedups ranging from 2x to 5x. In complex microcontroller examples, manual analysis often exceeds 25 minutes, while ConnChecker completes the task in under 8 minutes. This efficiency stems from its automated traversal and filtering of the dependency graph, which narrows down the root-cause candidates without requiring exhaustive manual inspection.

\begin{table}[H]
  \centering
  \footnotesize
  \setlength{\tabcolsep}{2pt}
  \renewcommand{\arraystretch}{1.1}
  \begin{tabular}{|>{\centering\arraybackslash}p{1.5cm}|p{2.5cm}|p{7.0cm}|>{\centering\arraybackslash}p{1.5cm}|>{\centering\arraybackslash}p{2.0cm}|>{\centering\arraybackslash}p{2.5cm}|}
    \hline
    \textbf{Connection} & \textbf{Design type} & \textbf{Connection's characteristic} & \textbf{No. of connection's edges} & \textbf{Manual debug time (HH:MM:SS)} & \textbf{Accelerated by ConnChecker (HH:MM:SS)} \\
    \hline \hline
    1  & Radar Sensor              & Combinational                                             & 7  & 00:01:40 & 00:01:00  \\
    2  & Radar Sensor              & Combinational                                             & 6  & 00:01:33 & 00:01:10  \\
    3  & Radar Sensor              & Combinational/Mixed-signal                                & 8  & 00:15:23 & 00:03:50  \\
    4  & Radar Sensor              & Combinational/Mixed-signal                                & 16 & 00:17:15 & 00:03:42  \\
    5  & Radar Sensor              & Combinational/Mixed-signal                                & 16 & 00:13:32 & 00:03:51  \\
    6  & Radar Sensor              & Combinational/Mixed-signal                                & 17 & 00:17:26 & 00:03:46  \\
    7  & ATV Microcontroller       & Combinational/Mixed-signal                                & 41 & 00:30:24 & 00:08:12  \\
    8  & ATV Microcontroller       & Combinational/Mixed-signal                                & 40 & 00:28:12 & 00:07:16  \\
    9  & ATV Microcontroller       & Combinational/Mixed-signal                                & 41 & 00:25:16 & 00:07:46  \\
    10 & ATV Microcontroller       & Combinational/Sequential/Mixed-signal                     & 45 & 00:22:14 & 00:06:32  \\
    11 & ATV Microcontroller       & Combinational/Sequential/Mixed-signal/Multi-clock domain  & 47 & 00:25:45 & 00:07:35  \\
    12 & ATV Microcontroller       & Combinational/Sequential/Mixed-signal/Multi-clock domain  & 47 & 00:23:40 & 00:06:52  \\
    \hline
  \end{tabular}
  \caption{Analysis Flow 2 - Cases with no structural connection.}
  \label{tab:flow-2-table}
\end{table}

\begin{figure} [H]
    \centering
    \includegraphics[width=0.6\linewidth]{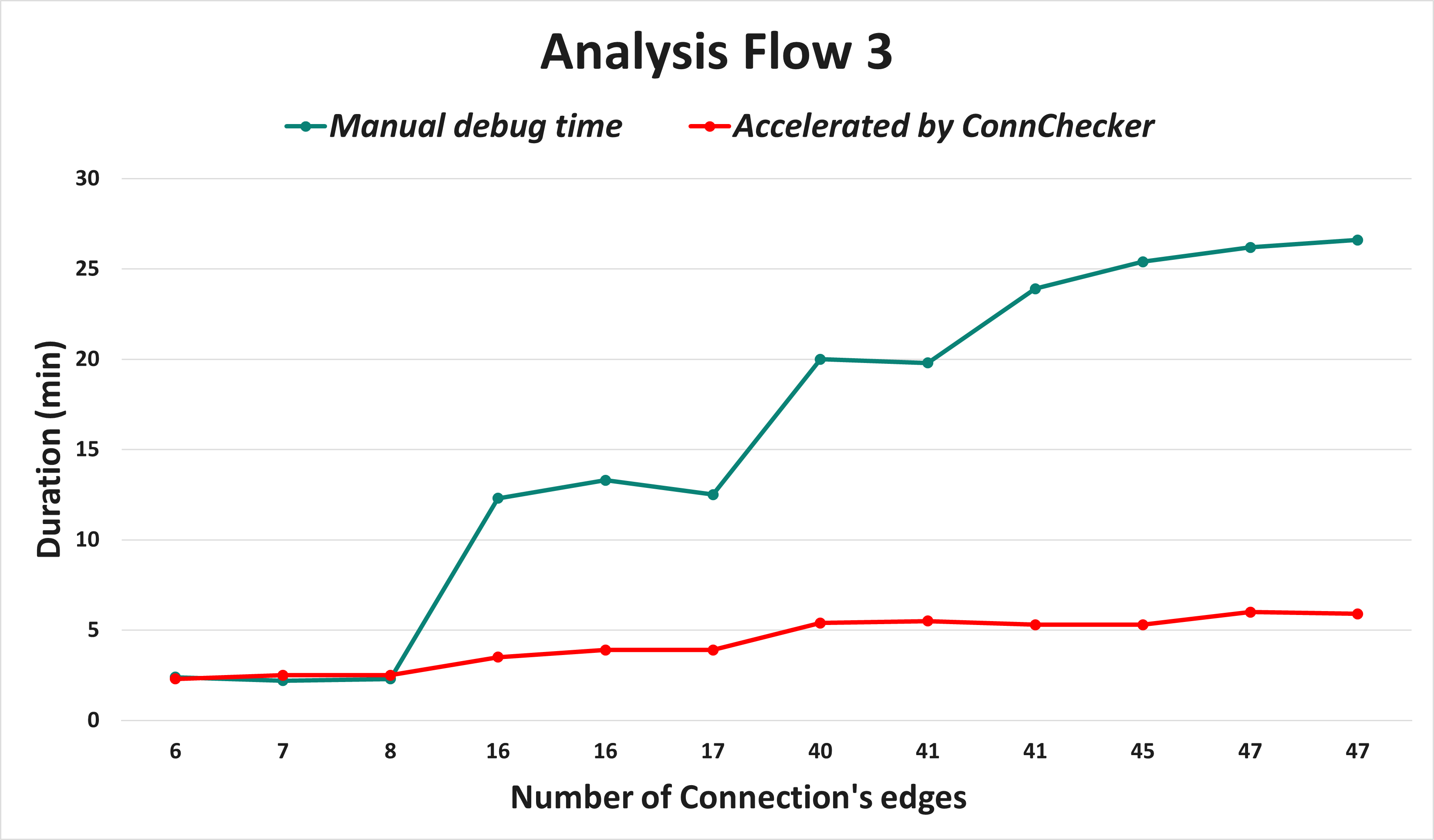}
    \caption{Graphical comparison of Table~\ref{tab:flow-3-table}.}
    \label{fig:flow-3-graph}
\end{figure}

Table~\ref{tab:flow-3-table} and Figure~\ref{fig:flow-3-graph} show results for Analysis Flow 3. This scenario is uncommon in radar sensor designs but frequently observed in automotive microcontrollers, where constraint complexity is significantly higher. For simple digital connections with fewer than five edges, manual debugging is nearly as fast, occasionally faster, due to the ease of schematic tracing. However, as connection complexity increases, ConnChecker shows clear advantages: achieving roughly 60\% time savings for sensor connections with 5-10 edges, and up to 80\% for microcontroller paths with 40–50 edges.

\begin{table}[H]
  \centering
  \footnotesize
  \setlength{\tabcolsep}{2pt}
  \renewcommand{\arraystretch}{1.1}
  \begin{tabular}{|>{\centering\arraybackslash}p{1.5cm}|p{2.5cm}|p{7.0cm}|>{\centering\arraybackslash}p{1.5cm}|>{\centering\arraybackslash}p{2.0cm}|>{\centering\arraybackslash}p{2.5cm}|}
    \hline
    \textbf{Connection} & \textbf{Design type} & \textbf{Connection's characteristic} & \textbf{No. of connection's edges} & \textbf{Manual debug time (HH:MM:SS)} & \textbf{Accelerated by ConnChecker (HH:MM:SS)}  \\
    \hline \hline
    1  & Radar Sensor              & Combinational                                             & 7  & 00:02:12 & 00:02:30  \\
    2  & Radar Sensor              & Combinational                                             & 6  & 00:02:24 & 00:02:18  \\
    3  & Radar Sensor              & Combinational/Mixed-signal                                & 8  & 00:02:20 & 00:02:28  \\
    4  & Radar Sensor              & Combinational/Mixed-signal                                & 16 & 00:12:17 & 00:03:32  \\
    5  & Radar Sensor              & Combinational/Mixed-signal                                & 16 & 00:13:16 & 00:03:56  \\
    6  & Radar Sensor              & Combinational/Mixed-signal                                & 17 & 00:12:27 & 00:03:54  \\
    7  & ATV Microcontroller       & Combinational/Mixed-signal                                & 41 & 00:19:47 & 00:05:32  \\
    8  & ATV Microcontroller       & Combinational/Mixed-signal                                & 40 & 00:20:01 & 00:05:26  \\
    9  & ATV Microcontroller       & Combinational/Mixed-signal                                & 41 & 00:23:56 & 00:05:16  \\
    10 & ATV Microcontroller       & Combinational/Sequential/Mixed-signal                     & 45 & 00:25:23 & 00:05:18  \\
    11 & ATV Microcontroller       & Combinational/Sequential/Mixed-signal/Multi-clock domain  & 47 & 00:26:14 & 00:05:59  \\
    12 & ATV Microcontroller       & Combinational/Sequential/Mixed-signal/Multi-clock domain  & 47 & 00:26:34 & 00:05:53  \\
    \hline
  \end{tabular}
  \caption{Analysis Flow 3 - Cases with only structural connection.}
  \label{tab:flow-3-table}
\end{table}

Overall, the graph-based framework was validated on two industrial designs, including one of Infineon’s largest microcontroller chips, confirming its applicability to large-scale systems. Although experiments were limited to the observed complexity, performance trends across three graphs (red lines) show runtime remains nearly constant as connection complexity increases. This demonstrates strong scalability and suggests the approach can extend to diverse design types.

\subsection{Discussion}

From our experimental evaluation, we identified three primary factors contributing to the significant increase in manual debug time: mixed-signal connectivity, multi-clock domain crossings, and multiple failure points along a single path. In this section, we present how ConnChecker’s graph-based workflow effectively addresses each of these challenges.

\subsubsection{\textbf{Mixed-Signal Connectivity}}

\begin{figure}[H]
    \centering
    \includegraphics[width=0.9\linewidth]{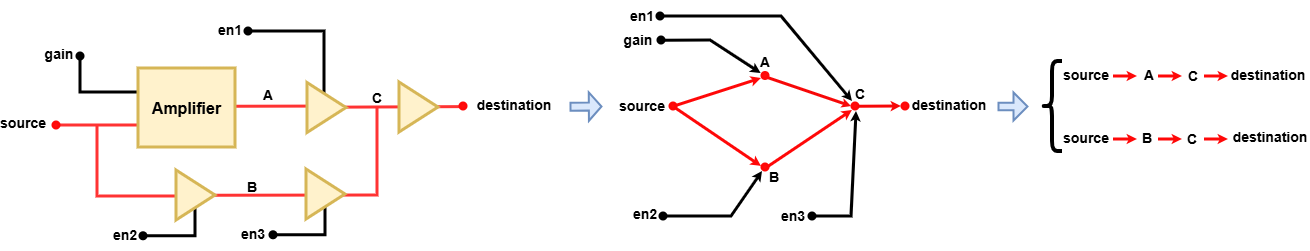}
    \caption{Example for multiple paths from source to destination exists}
    \label{fig:mixed-signal}
\end{figure}

Mixed-signal connectivity is a key factor contributing to the substantial increase in manual debugging time. In traditional digital connectivity debugging workflows, engineers can often proceed without deep knowledge of the DUT, as there is typically a single, well-defined source-to-destination path. Standard tools such as waveform and schematic viewers are usually sufficient to trace failures and identify root causes. However, mixed-signal connectivity introduces significant complexity. Multi-driver nets and feedback loops can result in multiple potential source-to-destination paths as shown in Figure~\ref{fig:mixed-signal}, making it difficult to identify the correct one without in-depth DUT expertise. Even after the correct path is determined, engineers must still trace the failure point, understand the enabling conditions, and implement a fix. These activities add further effort and time. ConnChecker helps make this process more manageable. When the correct path cannot be determined upfront, it applies its Analysis Flow across all candidate paths and presents the results to engineers for further inspection. These insights allow engineers to evaluate the characteristics of each connection and efficiently identify the correct path. This approach may not be exhaustive and may still require manual analysis, particularly in corner cases where a large number of candidate paths can lead to state-space explosion and increased formal tool runtime, but evaluation on two industrial designs found that the number of candidate paths per connection was consistently fewer than five. Overall, despite some limitations, ConnChecker still provides a structured methodology that improves traceability and supports more efficient debugging.

\subsubsection{\textbf{Multi-Clock Domain Crossings}}
\begin{figure}[!h]
    \centering
    \includegraphics[width=0.9\linewidth]{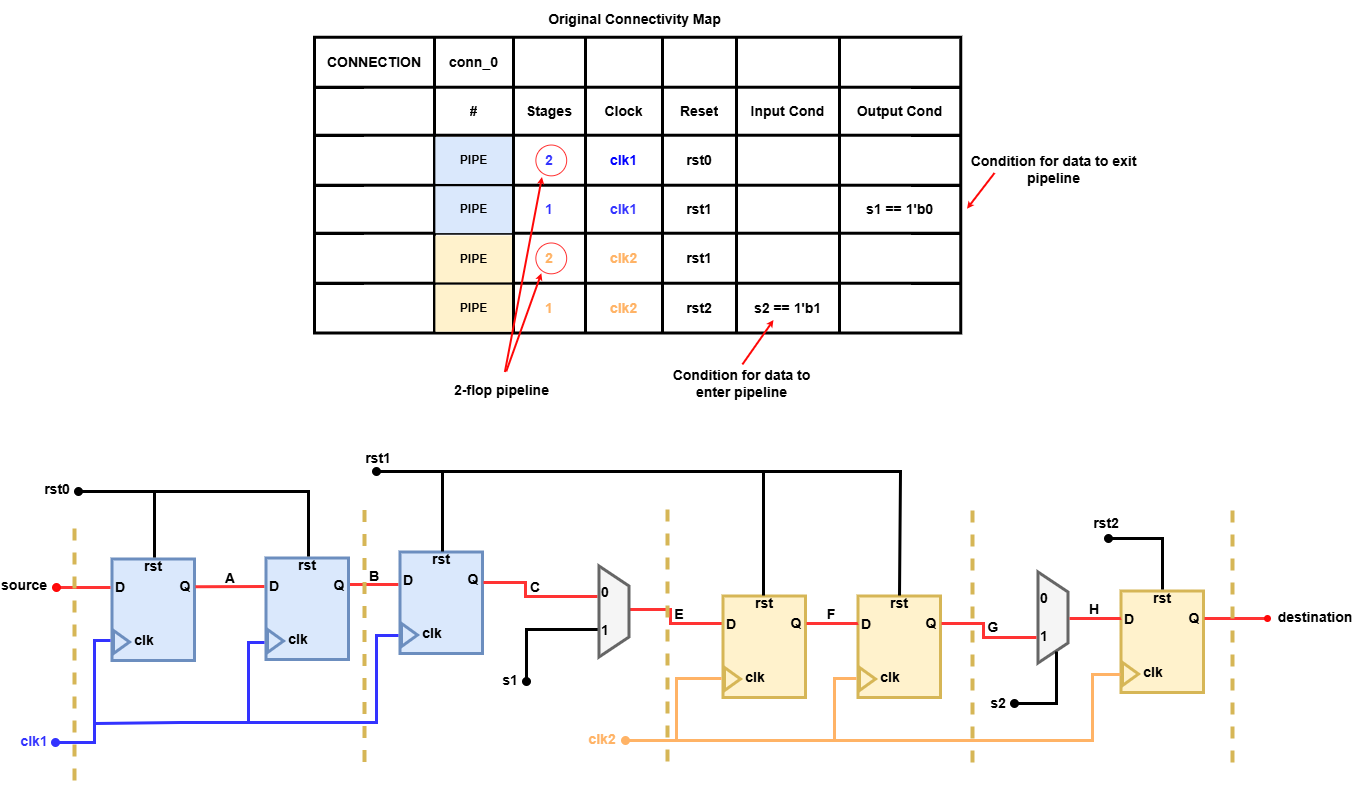}
    \caption{Example for multi-clock domain connection}
    \label{fig:PIPE}
\end{figure}

Multi-clock domain crossings primarily affect breakpoint localization and path reconstruction, rather than the identification of source-to-destination connectivity. As illustrated in Figure~\ref{fig:PIPE}, constructing a reliable connectivity map, which in this case follows the Jasper connectivity map format, requires detailed information about each PIPE domain, including associated clock and reset signals, the number of sequential elements, and the conditions governing domain transitions. In a graph-based framework, this process is formalized through segment-level analysis. Each segment explicitly captures clock control relationships for sequential elements, enabling systematic construction of the connectivity map and facilitating the identification of failure points. This representation improves visibility across domains and supports more efficient debugging in designs with complex clocking schemes.

\subsubsection{\textbf{Multiple Failure Points Along a Single Path}}

Paths with multiple failure points often require iterative debugging, where each issue is resolved sequentially across multiple tool runs. This process increases verification time and effort. ConnChecker reduces iteration by identifying all failing segments in a single pass, allowing engineers to address multiple issues concurrently.

Across these scenarios, a graph-based framework representation with segment-level reporting helps engineers focus on the right paths, expose all failing segments early, and reduce iterative rework, thereby shortening debug time and improving traceability. While extreme cases with many candidate paths can still stress ConnChecker, the approach consistently streamlines analysis and concentrates expert effort where it delivers the most value.

\section{Future Work} \label{future_work}

Despite our graph-based approach providing significant advantages, it faces certain limitations. A notable challenge occurs when multiple functional or structural paths exist between the source and destination node, as discussed earlier in the context of mixed-signal connections. Automatically identifying the correct path for analysis remains an unresolved problem and often requires input from designers. At present, the most practical solution is to supply engineers with comprehensive information that accelerates this decision-making process. Such information may include: the shortest path, the longest path, paths containing sequential elements, paths involving mixed-signal components, and the connectivity conditions associated with each path. Currently, ConnChecker processes all potential paths and reports them to the user for further debugging, rather than selecting a single definitive path.

Another limitation arises in scenarios involving bus connections. Unlike single-bit signals, a bus is represented as multiple individual connections, with the number of connections determined by its bit-width. For example, a 64-bit bus requires ConnChecker to process 64 separate dependency paths. This exhaustive approach is computationally expensive and highly inefficient. The key challenge is reducing the search space. One promising strategy is to leverage data values from counterexample (CEX) reports: by extracting the source and destination signal values and performing bitwise comparisons, then ConnChecker can focus only on mismatched bits rather than process every connection. Furthermore, even when mismatches are detected across the entire bus, prioritizing critical endpoints such as the least significant bit (LSB) and most significant bit (MSB) can further streamline analysis. This targeted analysis significantly minimizes overhead and accelerates debugging by isolating the actual points of failure.  

Although our evaluation was limited to the Cadence JasperGold Connectivity App, the ConnChecker methodology is broadly applicable to other formal connectivity tools. Since it operates on standard outputs, like dependency graphs and failure logs, processed independently via a Python-based system, minimal adaptation is required. We encourage users of alternative formal flows to explore this approach and share feedback through future industry forums.

Our main focuses now are on improving path selection heuristics in multi-path scenarios and optimizing bus-level analysis to reduce computational overhead. These extensions will build on ConnChecker’s graph-based foundation, further advancing its automation capabilities and applicability to increasingly complex verification environments.

\section{Conclusion} \label{conclusion}

ConnChecker presents a scalable, graph-based framework that automates root-cause analysis for formal connectivity verification using the Cadence JasperGold Connectivity App. By transforming tool outputs into structured workflows, it addresses a long-standing gap in debugging automation, which has remained largely manual despite standardized methodologies. Experimental results on industrial designs show an average 4x reduction in debug time, with even greater improvements in high-complexity cases. Although current limitations exist, like limited support for bus-level connections, ConnChecker fills a critical void in the verification ecosystem and establishes a foundation for future innovation in connectivity checking, contributing to faster debugging, improved scalability, and reduced time-to-market.

\printbibliography[heading=bibintoc]

\end{document}